# Highly oriented EuO nanocrystalline films via reduction process - NIR optical response

A. Mariscal,[1,*] A. Quesada,[2] A. Tarazaga Martín-Luengo,[3] Miguel A. García,[2,4] A. Bonanni,[3] J. F. Fernández,[2] R. Serna.[1]

[1]*Laser Processing Group, Instituto de Óptica, IO-CSIC, C/Serrano 121, Madrid, 28006, Spain*

[2]*Ceramics for Smart Systems Group, Instituto de Cerámica y Vidrio, C/ Kelsen 5, Madrid, 28049, Spain*

[3]*Institut für Halbleiter-und-Festkörperphysik, Johannes Kepler University, Altenbergerstr. 69, Linz, A-4040, Austria*

[4]*Instituto de Magnetismo Aplicado, UCM-ADIF-CSIC, Ctra. A6-km 22.5, Madrid, 28230, Spain*

**Abstract:**

Nanocrystalline textured EuO thin films are prepared by an oxygen loss process from a pure $Eu_2O_3$ bulk ceramic target through pulsed laser deposition in vacuum at room temperature. X-ray diffraction spectra evidence a well-defined diffraction peak corresponding to the EuO phase textured along the (110) direction. Analysis of the XRD peak profile indicates that the films are nanocrystalline (average crystallite size of 11 nm) with a compressive residual strain. The formation of stoichiometric EuO is further confirmed by a strong signal from $Eu^{2+}$ in the X-ray photoelectron spectra. The complex refractive index in the near infrared has been determined by spectroscopic ellipsometry and shows that the EuO films have a high transparency ($k < 10^{-3}$) and a refractive index of 2.1. A band-gap shift of 0.25 eV is found with respect to the EuO bulk. These films, deposited by an accessible and efficient method, open a new route to produce EuO films with optical quality, suitable for NIR optoelectronic components.

---

[*] Author to whom correspondence should be addressed. Electronic mail: antonio.mariscal@csic.es

## 1 - Introduction

The ferromagnetic semiconductor EuO is one of the most interesting compounds for spintronics devices [1,2]. The high spin polarization, magnetic properties [3,4], and the potential to integrate EuO with relevant semiconductors such as Si, GaN, and GaAs [1,5,6], makes this material appealing from basic and applied standpoints. However, the synthesis of EuO is reported to be remarkably challenging [7], which has hindered a more prolific research in a wider range of fields besides magnetism and spintronics. Indeed, the most stable oxide of Europium is $Eu_2O_3$ and, therefore, to obtain the reduced oxide EuO requires a careful control of the oxygen pressure during preparation. With the exception of chemical solid-state reactions [7,8], most of the preparation methods of thin EuO films, are based on the evaporation from a pure Eu bulk material in the presence of a carefully controlled oxygen environment. Using this methodology most of the report are on the growth by molecular beam epitaxy (MBE) [9,10], and, to a lesser degree, by other deposition techniques like *e.g.*, pulsed laser deposition (PLD) [11]. Among these techniques, PLD has the advantage of versatility to fabricate almost any material in thin film configuration [12], specially complex oxides [13–15], and film deposition can be achieved from vacuum to high gas pressure environments. Even if PLD is usually referred to as a technique that allows stoichiometric transfer from the target to the substrate, this stoichiometric transfer occurs only in the case of some oxides, such as $Al_2O_3$ [16]. For most oxides in vacuum, the ablation process produces an oxygen loss with respect to the original target, that can be restored controlling the background pressure [17–19]. In particular, target compounds consisting of heavy metals and light and volatile oxygen (e.g. $Eu_2O_3$) results in a strong oxygen deficiency due to the dissociation of the europium-oxygen bonds during the laser interaction with the target surface or afterward by collisions during plume propagation [20]. The dynamics of the plasma expansion are complex because there are many physical processes involved [21,22], however, this oxygen loss effect is an experimental well studied phenomenon that have been reported for a widely range of materials and works in thin films including, lead–niobium-germanate glass [23],

LiNbO$_3$ [24], ZnO$_{1-x}$ [19], SrTiO$_{2.5}$ [25], TiO$_x$ [26], Ga$_2$O$_x$ [27] and indium tin oxide [28] as some examples.

In the present work, we show how to profit from the known kinetic process that produces a deficiency of oxygen in the PLD deposited layers, to facilitate the preparation of EuO films by oxygen loss process from a Eu$_2$O$_3$ target in vacuum at room temperature. This procedure, eliminates the need of using controlled O$_2$ gas partial pressures as has been done up to date for the production of EuO in most reported works by physical deposition techniques [4,10,11,29]. We show that besides obtaining films with excellent EuO stoichiometry, they also exhibit an excellent crystalline texture, and notable magnetic and optical properties. The later are specially relevant because have been rarely studied in thin films [30], although they are relevant to achieve the integration of EuO in optoelectronic components.

## 2 - Methods

EuO thin films were prepared by PLD using an ultraviolet (UV) laser ArF excimer ($\lambda$ = 193 nm; 20 ns pulse duration). The laser beam is focused at an angle of incidence of approximately 45° onto the Eu$_2$O$_3$ target. Silicon wafers are employed as substrates and the ablation is performed in the off-axis configuration, *i.e.*, the center of the substrate is not aligned with the plasma expansion axis. The substrate is rotated during deposition in order to obtain large areas of homogeneous thickness. The films are prepared using an energy density of (3.5 ± 0.2) J cm$^{-2}$ at 20 Hz, with the substrates at room temperature (RT) and as a function of the background pressure from 10$^{-5}$ to 10$^{-4}$ Pa (see supplementary material). The thickness of the EuO films is ≈100 nm and without breaking the vacuum, a layer of 15 nm of amorphous Al$_2$O$_3$ are deposited on top in order to protect them from oxidation upon ambient exposure. For the fabrication of the films, a Eu$_2$O$_3$ monoclinic phase target (fabricated by Ceramics of Smart Systems Group – CSIC) [31] and a commercial Al$_2$O$_3$ target are used.

The stoichiometry of the EuO films has been assessed by x-ray photoelectron spectroscopy (XPS) measurements, using a Theta Probe XPS system from Themofisher with a monochromatic 1486.6 eV

Al-Kα X-rays operated at 15 kV and emission current of 6.7 mA. A maximum spot size of the X-ray beam of 400 μm in diameter has been employed, and a constant energy mode with constant pass energy of 50 eV (1.00 eV full width at a half maximum (FWHM) on Ag $3d_{5/2}$) has been used. In order to neutralize the charge build-up on the investigated surface, a standard dual flood gun provides simultaneously a beam of low energy electrons (2 eV) and a beam of low energy Ar-ions. The component of the C 1$s$ photoelectron peak at a binding energy of 284.8 eV corresponding to a C − C environment is taken as a binding energy reference. The background residual pressure of the high vacuum analysis chamber is held in the low range of $10^{-7}$ Pa. In order to ensure the integrity of the layer under study, only the capping layer has been sputtered with 3 KeV Ar$^+$ ions with a scan area of (2 x 2) mm$^2$ and a 200 seconds sputter time. The formation of crystalline phases is assessed by means of X-ray diffraction (XRD) (D8 Advance, Bruker, Germany) with Cu K$_α$ radiation. Spectroscopic ellipsometry (SE) measurements are performed in the 800 nm to 1700 nm wavelength range at incidence angles of 60°, 65° and 70° using a VASE ellipsometer (J.A. Woollam Co., Inc.) in order to determine the deposition rates, thickness and optical parameters of the material under investigation. The magnetic characterization has been carried out *via* a superconducting quantum interference device (SQUID) from Quantum Design.

**3 - Results and discussion**

*3.1 – Structural properties*

Several samples as a function of different pressures from $10^{-5}$ to $10^{-4}$ Pa in order to find the optimal vacuum pressure to obtain the EuO thin films. The XRD spectra show that when the vacuum pressure decreases to values of $10^{-5}$ Pa, the formation of the EuO nanocrystals increases (see supplementary material for further details). Therefore the results show how the vacuum pressure is a key parameter to control the stoichiometry of the deposited thin films. These findings are in agreement with those reported of the growth of $Eu_2O_3$ thin films in a previous work. In this work we reported that for pressures of $10^{-4}$ Pa, the europium oxide deposited films have a 3+ ($Eu_2O_3$) chemical state with some small traces of the 2+ oxidation state [32]. Conversely, in this work, we have achieved a 2+ oxidation state when we deposit at pressures of $10^{-5}$ Pa. These results will be further confirmed in the chemical characterization section. An example of the XRD spectrum of a EuO films grown at a vacuum pressure of $4.4·10^{-5}$ is presented in Fig. 1 and shows two high intensity peaks, the first at 33.03° corresponding to the (211) reflections from the Si substrate (01-072-1088), and the second centered at 50.49° and matching the position of the (220) orientation of EuO (00-018-0507). Since EuO crystallizes in a FCC lattice, only reflections with Miller indices all odd or even like ((111), (200), (220))… are permitted. Further analysis of the XRD spectrum provides in-depth information on the structure of the EuO film. The fact that the orientations (200) - intensity of 100% - and (100) - intensity of 90% - are not observed in the spectra points at a film strongly textured in the (110) direction. The significant broadening of the EuO (220) peak with respect to the Si one indicates that the film is formed by nanocrystallites. Furthermore, the EuO (220) peak shows a Lorentzian tail with a very good fitting as evidenced in Fig. 1, suggesting a slight contribution from non-uniform strain. In order to estimate the size of the nanocrystallites of EuO in the film we apply the Scherrer equation obtaining a mean size of the crystallites of ≈11 nm and concluding that the EuO film is nanocrystalline (more details in supplementary material).

Moreover, it is worth underlining that there is a significant shift of the (220) peak from 50.19° to larger angles, 50.49°, pointing at residual compressive uniform strain in the (110) plane. This result is in agreement with the high density films fabricated by PLD, and with the fact that the $Al_2O_3$ capping layer can also introduce residual strain in the europium oxide thin films [32]. It is likely, that the texture of the films is generated by the particular geometry of the growth configuration, combining off-axis and rotation. In the off-axis configuration the species reach the substrate with an oblique angle of incidence. Recently, it has been reported that under these non-conventional growth arrangement, it is possible to fabricate complex and textured nanostructures [33], compatible with the high quality textured nanocrystalline EuO films obtained in the present work.

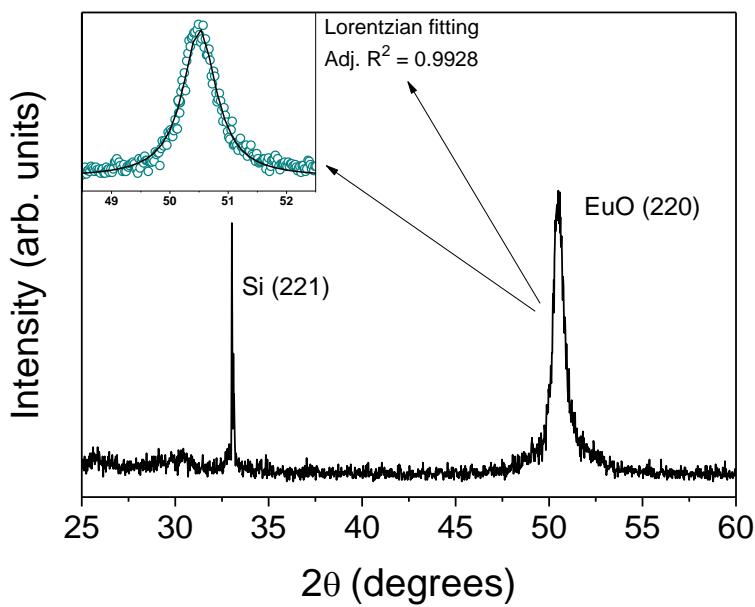

Fig. 1: XRD spectrum of EuO films with Lorentzian fitting for (220) EuO peak.

*3.2 – Chemical characterization*

In Fig. 2 (a) the Eu 3*d* high resolution XPS spectrum of a EuO film grown under $10^{-5}$ Pa is shown. The $Eu^{2+}$ $3d_{5/2}$ and $Eu^{2+}$ $3d_{3/2}$ spin-orbit components ($\Delta = 30 \pm 1$ eV) are the main features observed, in agreement with Eu in the 2+ oxidation state. The two main asymmetric peaks, centered at 1124.95 eV and 1154.14 eV respectively, are formed by a multiplet assigned to divalent Eu 3*d*, as theoretically calculated in previous works [29,34], together with a shake-up peak doublet (SU in the figure) at slightly higher energies with respect to the multiplet, and attributed to photoelectron energy loss common in rare earth compounds having unpaired electrons [35]. Two additional peaks with lower intensity and centered at 1134.03 eV and 1163.03 eV, are attributed to Eu in the 3+ oxidation state [36–38]. In this same position, satellite peaks were also reported [35]. In order to further support the assignment of these features, and to gain insight into those related to either satellites or weak contributions from $Eu^{3+}$, high resolution XPS spectra of Eu 4*d* have been collected, as reported in Fig. 2 (b). Here, no $Eu^{3+}$-related signal is detected, indicating that the low-intensity peaks in the Eu 3*d* high resolution spectrum at 1134.03 eV and 1163.03 eV can be ascribed to satellites and possibly to Eu$^{3+}$ at the very surface of the EuO layer, due to the fact that the 3$d$ photoelectrons have a smaller escape depth compared to the 4$d$ ones, providing then surface-sensitive information. It should be noted that the two components shown in the Fig. 2 (b) − both assigned to Eu$^{2+}$− are not related to a spin-orbit splitting, but to complex final state effects due to electron-electron interaction processes, associated with potential extra electronic transitions to empty 4f states [39].

This result shows that starting from a target with a Eu$_2$O$_3$ stoichiometry we have successfully formed films with a EuO stoichiometry, therefore we have achieved an oxygen loss process during PLD deposition. The oxygen loss process during the PLD has been reported in many works and it is related to the kinetics of the ablation plume [17–19,23–28]. For example, experiments using $^{18}$O as an ambient tracer gas have clearly shown that only 55% of the oxygen present in the films is directly transferred from the target; the remaining portion is supplied form the ambient gas [40]. During the laser ablation of a multi-element target, dense plasma at the target surface is formed in which all the elements of the compound in the form of neutral atoms and ions are included. Subsequently, the expansion of this plasma in the normal direction the target surface forms the so called "ablation plume". During the expansion of the plasma in vacuum (≤ 10$^{-4}$ Pa) multiple collisions of the species within the plasma occur. In a rough approximation they can be considered as elastic collisions [21,22], and therefore it is found that in each collision heavier ions suffer less dispersion than lighter ions. In our films Eu is a much heavier element (152 u) compared to O (16 u) and therefore O species suffer a larger dispersion. As a consequence a larger fraction of O species is deviated from the normal direction and are not able to reach the substrate, and therefore the O content in the substrate is reduced favoring thus the formation of the EuO film.

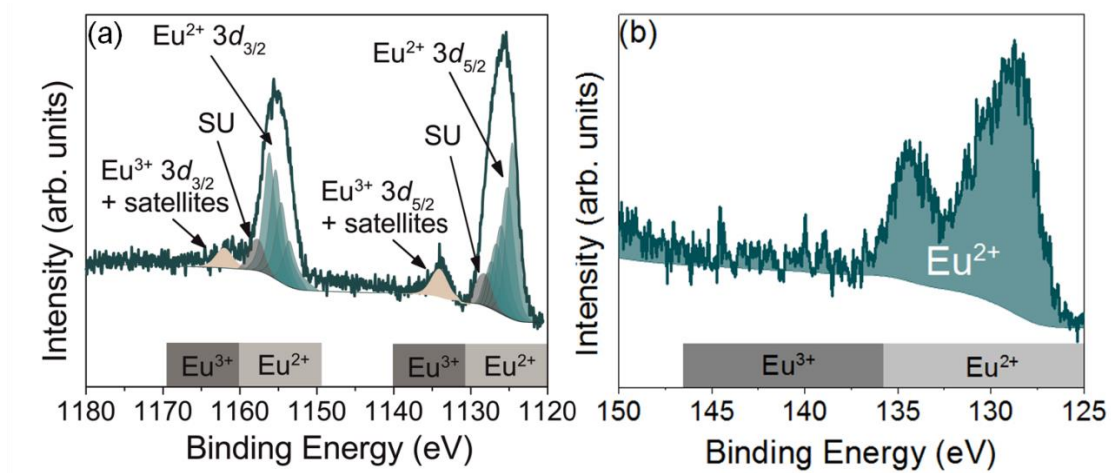

Fig. 2: High resolution XPS spectrum of EuO films. (a) Eu 3*d* (b) Eu *4d*.

*3.3 – Magnetization measurements*

Magnetization measurements (detailed information in supplementary material) show a maximum magnetic moment of 5.7 $\mu_B$/Eu calculated from the saturation magnetization, after subtracting the Si substrate diamagnetic component. This value is notable for a textured nanocrystalline thin film since it is comparable with the one reported for epitaxial PLD films [11], and even higher than some found for films grown by means of MBE [4].

*3.4 Near-infrared Optical response*

While the magnetic characterization of EuO is widely reported in literature, the information on its optical properties is scarce and the optical dielectric function is only available for bulk material in publications from the '70s [41,42]. Nevertheless, the optical constants of EuO thin films are prompting increased attention and recently, a study about the absorption coefficient of thin films of this material has been disclosed [30]. In this context, we have studied the complete optical response of the EuO textured layers, with the aim to deepen the understanding of their optoelectronic behavior that is of interest for potential semiconductor applications. Ellipsometric measurements have been carried out in the range between 600 and 1700 nm in order to obtain the optical parameters in the part of the visible region and in all near infrared (NIR), where the optical band-gap is expected [41,42]. In Figure 3 a) and b) the obtained values Ψ and Δ from the ellipsometric measurements at the angles 65°, 70° and 75° (with symbols) and the fittings (with solid black lines) are represented. In figure 3 c) the obtained *n* and *k* values in the NIR range for the EuO nanocrystalline thin films are plotted. The complex dielectric function has been modeled by a Cauchy formula in agreement with a fully transparent dielectric material in the NIR range. The refractive index is 2.1 and the absorption is negligible with *k* values of the order of $10^{-3}$ above 800 nm. These results allow us to conclude that the films show no band-gap in the NIR spectrum. This is a remarkable result, since the bulk EuO band-gap has been reported at 1.12 eV (1107 nm) [43], and therefore we would have expected a high absorption above this value.

If we look at the VIS region, below 900 nm, we observe the start of a clear absorption band with the k values increasing steadily. This band is attributed to the 4f to 5d transition of the semiconductor EuO [41]. To fit the optical data we have used the Tauc-Lorentz model. The Tauc-Lorentz model has been proven to be useful for modeling (with perfect Kramer-Kronig consistency) semimetals [44,45], semiconductors and dielectrics, in particular oxides [46,47]. It has shown to be specially successful for the modeling of the response of amorphous and nanocrystalline materials [48,49]. In our case, the Tauc-Lorentz method fits perfectly the ellipsometric measurements values as shown in Figure 3 a) b). From the fit of these values to a Tauc-Lorentz function we obtain a band-gap value of 1.37 eV which means that the band-gap has been shifted to shorter wavelengths by 0.25 eV, thus extending the transparency region to the NIR. A blue shift has already been reported for EuO nanolayers in which the shift was observed as the film thickness was reduced (1.1 nm – 1.4 eV) and this was related to quantum confinement in one dimension [30]. For the present films we do not have such a confinement because the films are over a 100 nm thick. However, the films are formed by nanocrystallites with average size of 11 nm, this means that the nanocrystals show a 3 dimensional quantum confinement effect and therefore a stronger quantum effect that can be the reason for the observed blue shift of the band gap. In addition, the role of compressive uniform strain cannot be neglected since it has shown to be the responsible of significant changes in the band-gap of semiconductor materials [50,51], and even there is a theoretical study of this effect for EuO [52]. The possibility of tuning the band-gap in EuO film has interesting implications, since this tunability will ease the integration of EuO, particularly with relevant direct semiconductors –for example GaAs- pointing to inject spin-polarized electrons for spintronics LEDs devices.

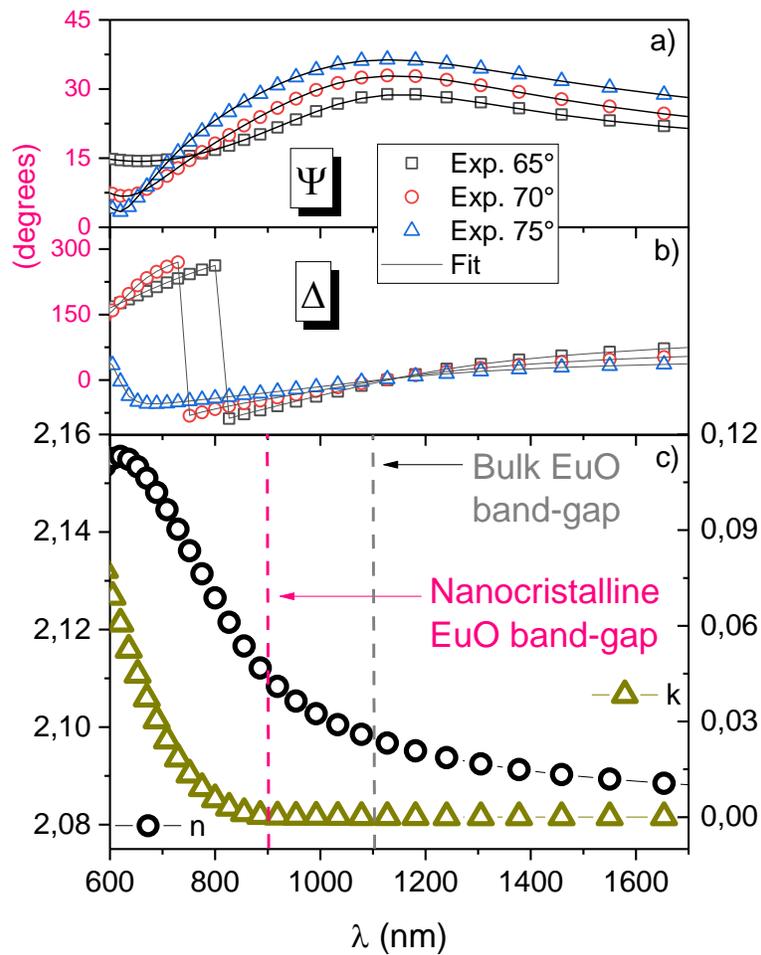

Fig. 3: a) and b), ellipsometric measurements (Ψ and Δ respectively) obtained at the angles 65°, 70° and 75°. The symbols represent the data and the solid lines the fittings. c) Refractive index, *n* and *k* values, in the VIS-NIR range for the EuO nanocrystalline thin films.

## 4 – Conclusions

In summary, we have devised a method to synthesize highly oriented nanocrystalline EuO thin films grown by an efficient oxygen loss process in vacuum at RT, and they have been investigated by means of XPS measurements to confirm the EuO stoichiometry. A textured structure of the layers – with strained nanocrystals of a size of 11 nm– is revealed by XRD profile analysis.

Concerning the optical dielectric function of the semiconductor EuO, the films are transparent (k< $10^{-3}$) in the NIR (905 nm to 1700 nm) with high refractive index of 2.1. The band-gap of the EuO nanocrystalline films is blue shifted from 1107 nm to 905 nm (1.12 eV to 1.37 eV) respect to the bulk, suggesting thus that the optoelectronics properties are potentially tunable either through quantum confinement and/or strain. This result paves the way for the accessible fabrication of tailored EuO films for integrate -optoelectronics and spintronics- devices.

## 5 – Supplementary material

See supplementary material for the details of the samples deposition as a function of the vacuum pressure, XRD spectra, calculation of the nanocrystalline size by Scherrer equation and the magnetic response characterization.

## 6 - Acknowledgements

This work has been financially supported by the Spanish Ministry of Economy and Competitiveness through the projects MINECO/FEDER TEC2015-69916-C2-1-R, MAT2013-47878-C2-1-R and MAT2013-48009-C4-1-P co-funded with FEDER funds. Support by the EC's Horizon2020 Research and Innovation Programme (grant 645776), by the Austrian Science Fund – FWF (P24471 and P26830), and by the NATO Science for Peace Programme (grant 984735) is acknowledged. A.M. acknowledges the financial support through BES-2013-062593.

# 7 - References

# Highly oriented EuO nanocrystalline films via reduction process - NIR optical response

A. Mariscal,[1,2] A. Quesada,[2] A. Tarazaga Martín-Luengo,[3] Miguel A. García,[2,4] A. Bonanni,[3] J. F. Fernández,[2] R. Serna.[1]

[1]*Laser Processing Group, Instituto de Óptica, IO-CSIC, C/Serrano 121, Madrid, 28006, Spain*

[2]*Ceramics for Smart Systems Group, Instituto de Cerámica y Vidrio, C/ Kelsen 5, Madrid, 28049, Spain*

[3]*Institut für Halbleiter-und-Festkörperphysik, Johannes Kepler University, Altenbergerstr. 69, Linz, A-4040, Austria*

[4]*Instituto de Magnetismo Aplicado, UCM-ADIF-CSIC, Ctra. A6-km 22.5, Madrid, 28230, Spain*

## Supplementary material

**Samples**

After the results obtained in the reference [1] where we found a small contribution of the europium oxide in the 2+ oxidation state, several samples (around 20) were made as a function of the, laser energy, temperature and vacuum pressure in order to understand how the growth parameter affects the formation of the europium monoxide. We have identified that the background vacuum was the key factor to obtain the best samples. In table I is presented a set of 3 samples as a function of the background pressure.

| Name | Vacuum Pressure (Growth) [$10^{-5}$ Pa] | Thickness [nm] |
|---|---|---|
| 4.4 | 4.4 | 140 |
| 9.7 | 9.7 | 164 |
| 14 | 14 | 106 |

---

[2] Author to whom correspondence should be addressed. Electronic mail: antonio.mariscal@csic.es

*Table SI: Data for the as-deposited samples. The data corresponding to sample 4.4 are shown in the manuscript*

**XRD Analysis**

The results of the XRD measurements of the as-deposited samples are presented in figure 1. For the vacuum pressure corresponding to $14 \cdot 10^{-5}$ Pa (sample 14), the spectrum presents no features of the europium monoxide in good agreement with the reference [1]. However, in reference [1] we find some contributions of the europium 2+ oxidation state by XPS analysis. This result means that we have a fraction of europium monoxide contribution in amorphous state which is not detectable by XRD measurements. For the sample growth at $9.7 \cdot 10^{-5}$ Pa (sample 9.7) we can distingue two broad peaks (30° and 50.49°) and one very narrow centered in 33.03°. The two broad peaks correspond to the (111) and (220) orientations of the europium monoxide (EuO) and the narrow peak corresponds to the (211) reflection from the Silicon substrate. Finally, for the sample growth at $4.4 \cdot 10^{-5}$ Pa (sample 4.4) we find a peak of the (220) orientation of the EuO but no features in the (111) direction, that means a strong texture in the (220) orientation. Also note, that this XRD measurement for the sample 4.4 is a different measurement from the showed in the manuscript. In the manuscript, the peak of the Silicon substrate (211) appears since here not. The apparition of this narrow Silicon peak is an artifact and is related with the orientation of the sample that can be give signal from the edge of the substrate.

The results show how controlling the vacuum pressure we can favored the formation of the europium monoxide by ablating a europium sesquioxide target. As we explain in the manuscript, for most oxides in vacuum, the ablation process produces a stoichiometry reduction with respect to the original target, that can be restored controlling the background pressure [2,3]. This oxygen deficiency is related to the kinetics of the ablation plume, which plays a crucial role in the stoichiometry of the films. For high vacuum pressures ($\leq 10^{-4}$ Pa), it is reported that heavier ions suffer less dispersion than lighter ions, for example, oxygen [4–6].

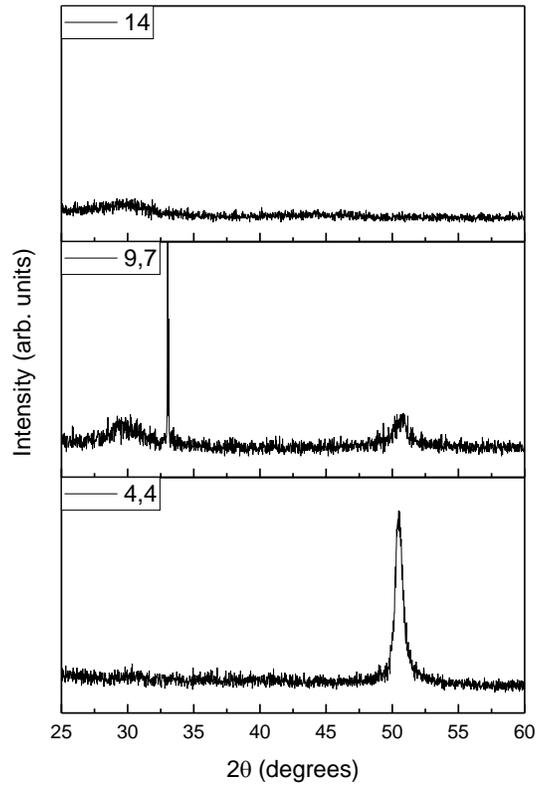

*Figure S1: XRD spectra for the samples 14, 9.7 and 4.4.*

**Scherrer analysis**

The characteristic size of the nanocrystallites in the film can be estimated using the Scherrer equation. The broadening of the diffracted beam that produces a parallel monochromatic radiation can be written as:

$$S = \frac{K\lambda}{(\beta_{obs} - \beta_{std})\cos\theta}$$

where $S$ is the mean size of the crystallite, $K$ a numerical constant related with the form of the crystalline and with a typical value of 0.93, $\lambda$ the wavelength of the incident X-rays, $\beta_{obs}$ the line broadening at FWHM of the relevant peak (in radians), $\beta_{std}$ the instrumental line broadening (in radians) and $\theta$ the Bragg angle.

Being $\beta_{std}$ instrument dependent, its quantification is not straightforward. In our case, we have used the (221) peak from the single crystal Si substrate at 33.03° as instrument line broadening. Inserting all the known values, as summarized in Table II, in the above equation we obtain a mean size of the crystallites of ≈11 nm for the sample 4.4 and we can conclude that the EuO film is highly textured and nanocrystalline.

| Peak pos. [°2Th] | β obs. [°2Th] | β std. [°2Th] | Crystallite size [Å] |
|---|---|---|---|
| 50.492 | 0.804 | 0.035 | 114 |

*Table SII. 4.4 sample parameters used for the Scherrer equation.*

**Magnetic response**

Figure S.1 shows the magnetization of the sample 4.4 as a function of the applied field (up to a maximum of 50 kOe) at RT (300 K) and 10 K. At 300 K, the films behave as a paramagnetic, as expected for EuO, whose Curie temperature ($T_c$) is reported to be about 70 K [7]. In contrast, the measurements at 10 K show a magnetization curve with virtually no hysteresis obtaining a coercivity of 0.5 Oe. In order to find out if the films are superparamagnetic, we perform a Langevin analysis on the magnetization curve realizing that it is not possible to fit the data to the Langevin function. Therefore, this result confirms the ferromagnetic nature of the EuO film. The most relevant consequence is that a maximum magnetic moment of 5.7 $\mu_B$/Eu is calculated from the saturation magnetization, after subtracting the Si substrate diamagnetic component. We notice that the 5 T applied field is not sufficient to fully saturate the sample, which could account for the slightly lower value compared to the theoretical moment of 7 $\mu_B$/Eu [8]. In addition, a reduction of the saturation

magnetization has been previously reported for magnetic oxide nanoparticles by size effects [9]. Concluding, the magnetic moment value obtained for the present EuO thin films (5.7 $\mu_B$/Eu), is comparable with the one reported for epitaxial PLD films [10], and even higher than some found for films grown by MBE [11].

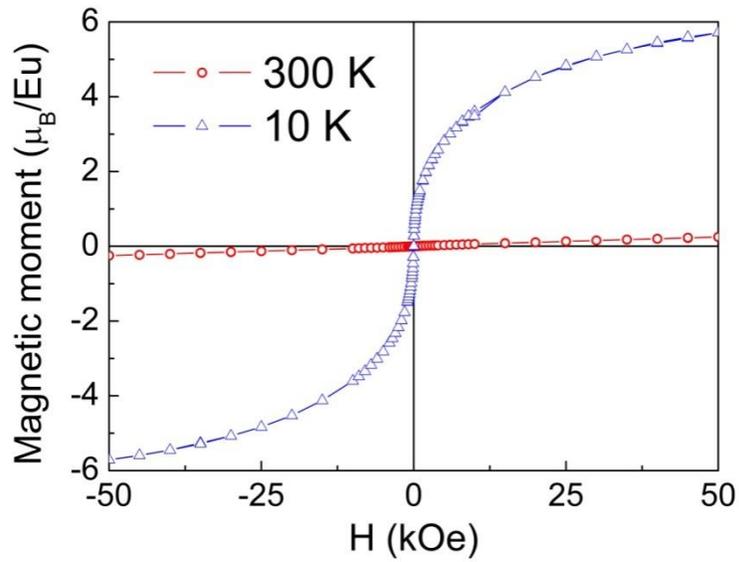

*Figure S2 : Magnetization as function of applied field at 300 K and 10 K*